# Disparate energy scaling of trajectory-dependent electronic excitations for slow protons and He ions

S. Lohmann[1] and D. Primetzhofer

Department of Physics and Astronomy, Uppsala University, Box 516, 751 20 Uppsala, Sweden

**Abstract**

We have simultaneously measured angular distributions and electronic energy loss of helium ions and protons directly transmitted through self-supporting, single-crystalline silicon foils. We have compared the energy loss along channelled and random trajectories for incident ion energies between 50 keV and 200 keV. For all studied cases the energy loss in channelling geometry is found lower than in random geometry. In the case of protons, this difference increases with initial ion energy. This behaviour can be explained by the increasing contribution of excitations of core electrons, which are more likely to happen at small impact parameters accessible only in random geometry. For helium ions we observe a reverse trend – a decrease of the difference between channelled and random energy loss for increasing ion energy. Due to the inefficiency of core-electron excitations even at small impact parameters at such low energies, another mechanism has to be the cause for the observed difference. We provide evidence that the observation originates from reionisation events induced by close collisions of helium ions occurring only along random trajectories.

---

[1] Corresponding author
  E-mail address: svenja.lohmann@physics.uu.se

# Article

Accurate knowledge of the energy deposition by energetic charged particles in matter is imperative for understanding astrophysical phenomena [1,2] and materials modification in extreme environments [3,4]. Knowing the average energy loss per path length provides the basis for analytical techniques such as ion beam analysis, while understanding of subsequent electronically driven processes allows for controlled tailoring of material properties by ion irradiation, implantation and sputtering, all of which are extensively used in materials research and widely spread in industrial applications [5]. Also, in hadron therapy, the induced energy spectrum of secondary electrons is decisive for a deepened understanding and accurate modelling of nanodosimetry [6].

The common concept for the description of energy dissipation defines the specific energy loss per unit path length as the stopping power $S$ of a material. $S$ is commonly considered a mean value along the trajectory of the ion, and several theoretical concepts have abandoned the complication of an internal structure of the target material and still successfully predicted $S$ [7,8]. However, even though we know that the transfer of energy from the moving ion to target constituents is impact parameter dependent [9], the underlying causes, i.e. the exact nature of how energy is transferred, are insufficiently understood, in particular for slow ions. Advanced theories predicting equilibrium (density functional theory DFT) and non-equilibrium conditions in solids (time dependent density functional theory TD-DFT) [10] are promising tools to accurately predict energy deposition even at the nanoscale, but still require to be benchmarked against suitable experimental data.

Experimental studies of energy loss, however, most often make use of the definition of *S*, and measure an effective average of the specific energy transfers experienced in the individual interactions along a trajectory, often in amorphous or polycrystalline targets. The capability of specifically studying the individual underlying processes in a comparison of experimental data to theory is, therefore, limited. As an example, predictions of energy loss due to binary collisions between the electrons and the ion using DFT typically match the experiment only for protons [11], whereas for heavier ions, due to contributions of more complex, local interactions [12], the observed energy loss is underestimated [13].

By predefining the impact parameters accessed in a specific experiment, a more detailed understanding of electronic excitation channels accessible to an energetic ion can be obtained. A selection of impact parameters can be achieved by employing materials with long-range order, i.e. single crystalline materials. When an ion enters a crystal with its direction of motion closely aligned with a major crystal axis, it will typically undergo only small-angle collisions with the nuclei. These collisions steer the ion through the channel formed by the strings and planes of the crystal, and it moves in an oscillating fashion at large impact parameters [14].

The electronic stopping power of fast light ions in single-crystalline materials has been studied by several groups [15–21]. At energies between 500 keV and several MeV, significant differences between channelled and non-channelled trajectories have been reported in silicon targets. In these studies, different experimental approaches like backscattering or transmission with either relatively thick samples or samples with an additional backing layer have been performed. At similar energies more recent studies with ultra-thin, self-supporting silicon foils have focussed on the angular distribution of transmitted protons [22–24].

For slower ions with energies in the medium energy regime (several ten to few 100 keV) very little systematic analysis exists. Recently, Wang et al. have constructed a transmission channelling set-up for a He ion microscope, but no measurements of electronic stopping have been performed [25]. The response from graphene to the impact of slow, highly charged ions has been investigated illustrating the significant influence of charge exchange processes on secondary electron emission in this system [26,27], but no impact parameter dependence of electronic energy deposition has yet been established in such experiments.

At these low energies, the character of ion-solid interaction changes and non-adiabatic processes become important. Thus, in this regime, impact parameter dependent phenomena will behave differently than observed at energies of several hundreds of keV/nucleon. Specifically, core-electron excitation is not as efficient, the structure of the density of states of the target becomes important and dynamic processes such as charge exchange [28] and formation of molecular orbitals [29] are expected to significantly contribute and alter the electronic response at the nanoscale, both in terms of number and energy of secondary electrons. Recently, significant advances in theoretical description of these dynamical processes in solids have been made due to the development of TD-DFT. Similar as for experiments, also from a theoretical point of view, the order of well-defined lattice locations and the periodicity of crystals make them easier to model than amorphous materials. Under these conditions, TD-DFT has been successfully employed to model transmission of ions through crystalline samples and results on multiple ion-target combinations have been published, for example H in Si and Ge [30] [31], Si in Si [32], Ni in Ni [33] and H in graphite [34].

In order to investigate the described dynamics in ion-electron interaction, we simultaneously record angular distributions and energy loss of helium ions and protons transmitted through thin, self-supporting, single-crystalline silicon foils. For the first time we compare helium and proton data at low energies below and around the Bragg peak, respectively. We show that, also for the lowest investigated energies, the energy dissipated along channelled trajectories is significantly less than along random trajectories. As a key result, we find the energy dependence and magnitude of this effect to fundamentally differ between protons and helium ions.

Experiments were performed with the time-of-flight medium energy ion scattering (ToF-MEIS) set-up at Uppsala University [35]. Helium ions and protons were detected with a large position-sensitive microchannel plate detector after transmission through thin, self-supporting Si(100) foils (Norcada Inc. "UberFlat" silicone membranes). Different foil thicknesses were employed to check for possible effects trivially related to trajectory length. The detector was positioned with its surface orthogonal to the initial beam direction, 290 mm behind the sample. Initial ion energies ranged between 50 keV and 200 keV. The ion energy after transmission through the sample was determined via its time-of-flight simultaneously with a position signal. Rutherford backscattering spectrometry was used to independently confirm thicknesses and purity of the samples.

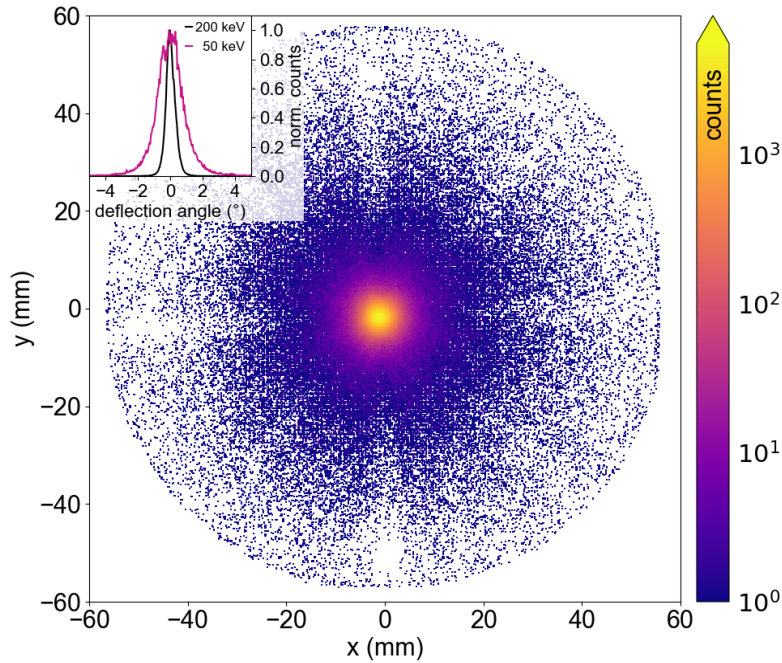

Figure 1: Spatial distribution of He ions transmitted through a 200 nm self-supporting Si(100) foil. The initial ion energy is 200 keV and the centre of the detector is positioned at 0° with respect to the initial beam direction. The main crystal axis is aligned with the beam direction. The inset shows the angular distribution of ions detected along a line parallel to the x-axis through the beam centre position (y = −2) for initial energies of 200 keV (black line) and 50 keV (magenta line).

Figure 1 shows the spatial distribution of $^4$He$^+$ ions with initial energies of 200 keV after transmission through a 200 nm Si(100) foil. The sample is aligned in such a way that the principal [100] crystal axis is parallel to the incident beam. In this geometry, the vast majority of ions is detected at small scattering angles axially symmetric around the incident beam direction. These ions, thus, travel through the crystal along a channelling trajectory. A minority of ions escapes the channels and reach the detector under larger scattering angles (note that the whole diameter of the detector corresponds to deflection angles ±11.5°). Some of these trajectories are, however, subject to the blocking effect, which results in reduced intensity at the projections of the crystal axes on the detector. In this way, a real-space image of the crystal structure becomes visible, with information on the particle energy simultaneously available for every pixel.

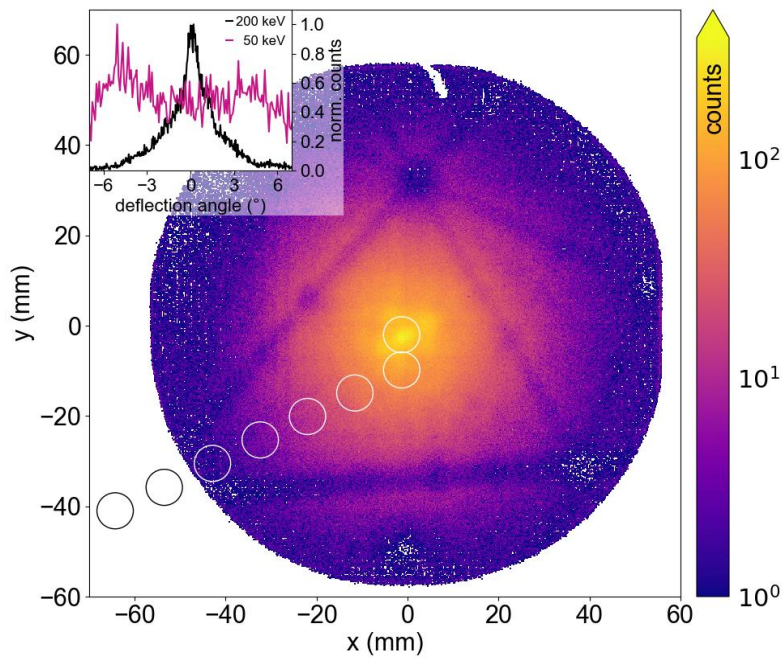

Figure 2: The sample is turned by 12° around the y- and by 7.5° around the x-axis compared to the channelling alignment used in Fig. 1. Otherwise, experimental conditions were identical. The circles show the position of the incident beams with respect to the crystal axes (visible by the blocking pattern) used for a detailed analysis of the specific energy loss (see text for details). The inset shows again the angular distribution of detected ions for initial projectile energies of 200 keV and 50 keV.

For Figure 2 we have turned the sample by 12° around the y-axis and by 7.5° around the x-axis. Otherwise, experimental conditions were identical as for Fig. 1. The beam is not aligned with one of the low-index crystal axes, therefore, we call this geometry "random". Again, the highest intensity can be observed in the centre of the detector, i.e. in the direction of the primary beam (denoted by the topmost circle in Fig. 2). The blocking pattern, i.e. the lines and nodes of reduced intensity, is readily visible. Note that the white region on the top is an artefact caused by a cable shadowing the detector. Both graphs also hold an inset showing the angular distribution of the recorded intensity along the x-axis. For decreasing energies, the angular spread increases, and the distribution for 50 keV He shown in the inset in Fig. 2 clearly illustrates how initially deflected primary ions can be subsequently channelled. Also, the observed angular spread is found lower for protons as expected (not shown). Generally, distributions of ions of different species and/or initial kinetic energy look qualitatively similar on the detector.

The present data permit to unambiguously correlate angular deflection and crystal orientation with the associated energy loss distribution. In the following, we assess the energy loss of ions for different, specific trajectories. For this aim, we varied the beam-crystal alignment successively between the channelled and the random geometry and selected ion trajectories ending in small, circular regions of interest on the detector (radius: 4 mm, scattering angle: ± 0.8°) around the projected position of the incident beam. We found no difference in energy loss for smaller regions of interest, i.e. for angular deflections small compared to the half-width of the observed channelling phenomenon. The circles in Fig. 2 visualise how the incident beam is positioned with respect to the crystal axes, as made visible by the blocking pattern, in the respective measurement. Note that in practice we rotated the crystalline sample; the detector was not moved, and the evaluated region of interest was kept at a fixed position.

Figure 3 shows the obtained energy loss spectra, i.e. the difference between initial ion energy and final energy, of He ions in Si(100) for different alignments between beam and crystal axis. In this example, the initial ion energy is 50 keV and the sample thickness is 200 nm. The figure legend shows the degree of crystal rotation away from the (100) channelling geometry ($\theta_x$ and $\theta_y$ denotes the rotation around the x- and the y-axis, respectively); an effect that is also visualised by the circles drawn into Fig. 2. All curves are normalised to their maximum value to ease comparison.

It is readily visible that the ions lose significantly less energy along channelled than along random trajectories. The broadening of the curves observed in the transition from channelling to random geometry is expected to be caused by an inclusion of qualitatively rather different trajectory types, in analogy to the skimming effect described for backscattering [36].

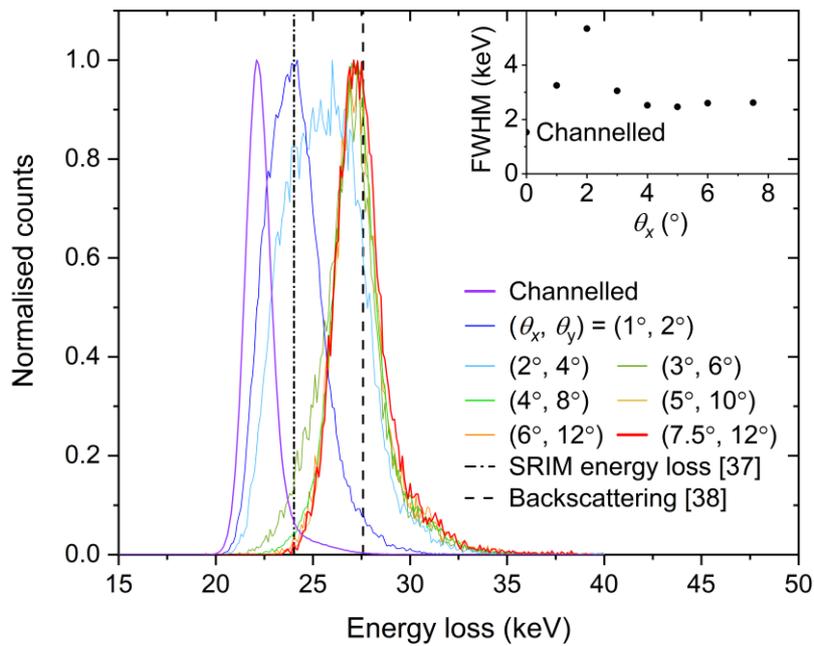

Figure 3: Energy loss spectra of He ions transmitted through a self-supporting, 200 nm thick Si(100) foil. The initial ion energy is 50 keV and data are taken from the same circular region of interest with 4 mm radius for each measurement. All curves are normalised to their maximum value. Vertical lines indicate the energy loss in amorphous Si from [37] and [38]. All other value pairs in the legend give the rotation around the x- and the y-axis with respect to this alignment. The inset shows the observed full-width at half-maximum of the energy loss distributions.

To quantify this phenomenon and to allow for a comparison between different energies and ion species, we study the energy loss in channelling geometry ($\Delta E_{ch}$) as a fraction of the random energy loss ($\Delta E_r$). Figure 4 shows the ratio $\Delta E_{ch}/\Delta E_r$ for He ions and protons with initial energies between 50 keV and 200 keV. The error bars take into account the uncertainty of the relative sample-detector distance caused by the finite size of the beam spot and the evaluated region of interest as well as the error in determining the flight time. The latter increases for shorter flight times, i.e. faster ions. The fraction $\Delta E_{ch}/\Delta E_r$ is found smaller than unity for all data points meaning that the energy loss along channelled trajectories is smaller than along random ones in all studied cases. The strength of this effect, however, strongly depends on ion species and initial ion energy. The energy loss of channelled protons at the lowest energies studied reaches about 0.95 of the random energy loss. With increasing energy, the fraction monotonously decreases, i.e. we observe a stronger difference between channelled and random geometries. For helium ions, we see a reverse trend: the channelled energy loss only reaches 0.82 of the random one at 50 keV but increases to about 0.9 at 200 keV.

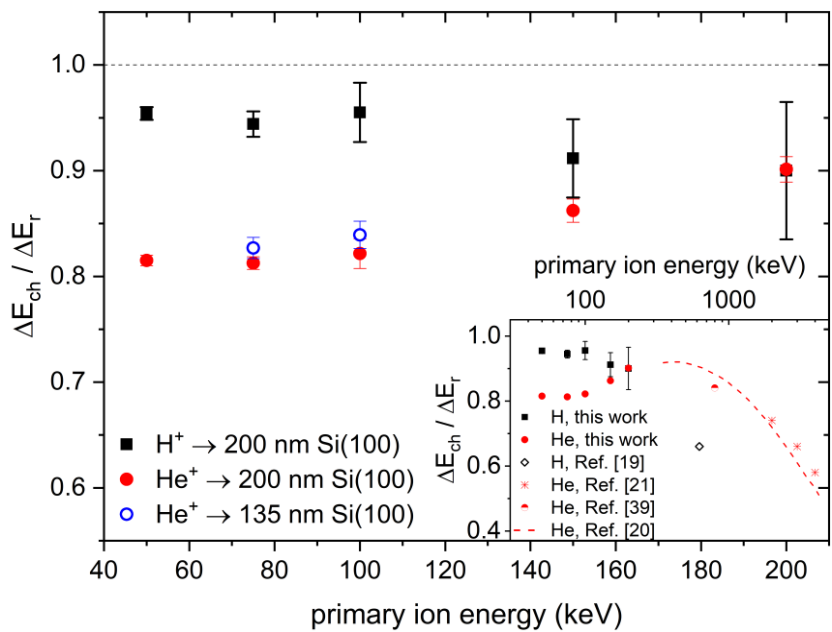

Figure 4: Energy loss along random ($\Delta E_r$) and channelled trajectories ($\Delta E_{ch}$) expressed in the fraction $\Delta E_{ch}/\Delta E_r$ for protons (black squares) and for He$^+$ ions (red full circles/blue open circles) transmitted through a Si(100) foil. The inset shows our data in comparison with energy loss ratios measured in Si(100) from the literature (H$^+$ data point from Ref. [19] (open diamond), He$^+$ data from Ref. [21] (red stars) and [39] (red half-filled circle). The red dashed line corresponds to a guide to the eye through measured data from Ref. [20].

For understanding these results it is necessary to consider the total energy loss as a sum of different processes [40]. The first contribution stems from the energy transferred to the target valence electrons via binary Coulomb collisions. These electrons can be considered as almost homogeneously distributed inside the target and even within the channel [18,39,41] and, thus, are expected to cause only a marginal difference between random and channelled geometries. Secondly, at sufficiently high energy transfers in ion-electron collisions, also highly localised silicon core electrons may be excited. Energy loss to these electrons is, therefore, expected to be impact parameter dependent and suppressed for channelled ions, which are steered towards the channel centre and experience no close collisions [39]. These effects in combination explain the observed decrease of the ratio $\Delta E_{ch}/\Delta E_r$ in previous studies at high energies and have been extensively discussed in literature [42]. The inset of Fig. 4 shows our results together with datapoints and an interpolation of data discussed in Ref. [20] for $\Delta E_{ch}/\Delta E_r$ at high energies. In this context, the observed energy dependence of our data for protons, as well as its extrapolation to higher energies, can be explained by the increasing efficiency of core-electron excitations and its consequent dominance over valence electron excitations [30].

In contrast, for helium ions with energies below 200 keV, the excitation of silicon core electrons becomes even more unlikely than for protons for two reasons. First, the maximum energy transfer in binary electron-ion collisions is much smaller than the binding energies of inner shells. Second, due to the stronger nuclear interaction, typical interaction distances are even larger than for protons also for random geometries. At 200 keV primary energy, scattering by 1° corresponds e.g. to an impact parameter of 0.048 Å and 0.084 Å for protons and He ions, respectively, with expected large impact on the probed core electron density. Thus, the observed difference in energy loss at these low energies, which is further increasing with lower initial ion energy, i.e. exhibits a maximum between ca. 400 keV and 500 keV, has to be attributed to a different energy loss mechanism.

We propose these observations to be rooted in repeated electron capture and loss of He projectiles caused by dynamic processes in the atomic level structure not included in a classical free-electron gas framework. Whereas for protons the 1s-level is typically resonant with the band in the solid, for He a more complex behaviour dependent on interaction time and impact parameters can be expected [43,44]. Generally, He ions of initial energies of several ten keV/nucleon capture an electron via Auger processes when they enter a solid at any impact parameter [45]. For neutral particles, excited states become relevant for the interaction [46], and a series of ionisation processes such as autoionisation from an excited state or collision induced reionisation, several of which show clear impact parameter dependence, exists. Also the formation of molecular orbitals and consecutive electron promotion has been shown to be active particularly at lower ion energies as sufficient interaction time is required for these non-adiabatic processes to occur [47]. In a global perspective, these processes will introduce a trajectory dependence of the mean charge state at low energies and thus impact electronic stopping.

In the following, we present a qualitative analysis based on these processes allowing to understand the increase of $\Delta E_{ch}/\Delta E_r$ at low energies based on cyclic capture and loss events affecting directly and indirectly the observed energy loss. The reionisation threshold for initially neutral He projectiles in large-angle collisions from Si is known from low energy ion scattering experiments [28] and found between 300 eV and 400 eV [28,48]. This energy corresponds to a distance of closest approach of about 0.35 Å for charge exchange during a scattering event, which can only be reached at scattering angles > 0.45° for 50 keV He projectiles. From the measured angular distribution of ions (conf. inset Fig. 1) we can deduce that a large fraction of channelled helium could indeed not have undergone charge exchange events, i.e. they are expected to have been neutral for a large part of their trajectory through the target. For a random trajectory the possible number of charge exchange events can be estimated geometrically. The probability to impact a circle with radius 0.35 Å around a silicon nucleus is about 0.04 per monolayer and thus would account to around 40 such close encounters for the whole sample (200 nm ≈ 1000 monolayers). In these processes a characteristic energy transfer of about 20 eV is observed [49]. While these direct losses have to be expected, they are clearly insufficient to explain the observed difference in energy loss. However, another consequence of these processes, including autoionisation due to formation of molecular orbitals, is that the probability that initially neutral He travels through the target in an ionised state is significantly enhanced for non-channelling geometries. Assuming a mean Auger neutralisation rate similar to that of Al, which is about $5.8 \times 10^{14}$ s$^{-1}$ [50], a typical neutralisation length of about 23 Å can be expected, effectively introducing a significant difference in the mean charge state for different trajectories. Since the electronic stopping power of the helium ion is significantly higher than that of the neutral atom [51,52], these ionisation processes can contribute to the observed differences in the energy loss of ion along channelled and random ion trajectories. On top of that, an energy-dependent charge state ratio will affect energy loss via modification of the interaction potential, as modelled for highly charged ions [53]. For the present system, this process will induce an additional energy dependent energy loss contribution,

The practical, consequences of the described scenario at the nanoscale are that ionisation densities of He ions, and other heavier species, are expected to oscillate along typical trajectories in amorphous or polycrystalline solids. This behaviour, apart from affecting the observed average energy loss per unit path length, would leave a clear signature in the secondary electron spectrum and subsequent electronically driven cascades. For accurate prediction of such decisive phenomena on the nanoscale, time-dependent modelling has to be improved by including charge exchange and by fully accounting for charge state dependency. Experimentally, measurements of exit charge states are required to obtain a more detailed understanding of these charge exchange mechanisms.

**Acknowledgements**

We would like to thank M. Motapothula for providing a 200 nm Si sample and for contributing to the very first stages of this project. Accelerator operation is supported by the Swedish Research Council VR-RFI (contracts #821-2012-5144 and #2017-00646_9) and the Swedish Foundation for Strategic Research (contract RIF14-0053).